\documentclass[twocolumn,prl,showpacs]{revtex4}

\begin{document}

\topmargin-1.0cm

\title {Nonuniqueness in spin-density-functional theory on lattices}
\author {C. A. Ullrich}
\affiliation
{Department of Physics and Astronomy, University of Missouri, Columbia, Missouri 65211}
\date{\today}
\begin{abstract}
In electronic many-particle systems, the
mapping between densities and spin magnetizations, $\{n(\bf r),{\bf m}(\bf r)\}$,
and potentials and magnetic fields, $\{v(\bf r),{\bf B }(\bf r)\}$, is known to be nonunique, which
has fundamental and practical implications for spin-density-functional
theory (SDFT). This paper studies the nonuniqueness (NU) in SDFT on arbitrary lattices.
Two new, non-trivial cases are discovered, here called {\em local saturation} and 
{\em global noncollinear} NU, and their properties are discussed and illustrated.
In the continuum limit, only some well-known special cases of NU survive.
\end{abstract}
\maketitle
Density-functional theory (DFT) \cite{HK1964,KS1965,Kohn1999}
is a widely used approach for calculating the electronic structure of atoms,
molecules, and many types of materials. The fundamental theorem of
Hohenberg and Kohn \cite{HK1964} establishes that the ground-state density $n({\bf r})$ of
a nonmagnetic, nondegenerate electronic system uniquely determines the scalar potential
$v({\bf r})$, apart from an arbitrary additive constant $C$. However,
many applications of interest involve electronic ground states that have a spin magnetization, ${\bf m}({\bf r})$, and/or 
are exposed to external magnetic fields, ${\bf B}({\bf r})$. Such situations can be handled with
spin-density-functional theory (SDFT) \cite{Barth1972,Capelle2001,Eschrig2001,argaman2002,gidopoulos,Kohn2004},
where the fundamental variable is the 4-density $\{n(\bf r),{\bf m}(\bf r)\}$, 
which couples to the 4-potential $\{v(\bf r),{\bf B }(\bf r)\}$.

Using the Rayleigh-Ritz variational principle, one can prove that a given physical
4-density uniquely determines the ground-state wave function $\Psi_0$ (apart from trivial phase factors).
This guarantees that any property of the system expressible in terms of $\Psi_0$ 
is a functional of the 4-density. Following the Hohenberg-Kohn theorem of DFT, one
might expect that there is also a unique map from ground-state wave functions to 4-potentials.
However, it was recognized long ago by von Barth and Hedin \cite{Barth1972}, and more recently by
Capelle and Vignale \cite{Capelle2001} and Esch\-rig and Pickett \cite{Eschrig2001} that such a unique 
correspondence does not exist. These authors showed that the extent of nonuniqueness (NU) in SDFT is much greater 
than the addition of a mere constant $C$ to $v({\bf r})$ for the nonmagnetic systems considered in DFT. 
As a consequence, some applications of SDFT such as the calculation of excitation energies or of one-electron 
spin gaps in half-metallic ferromagnets must be critically reexamined.

NU in SDFT means that an $N$-particle ground state 
$\Psi_0$ remains unchanged under addition of a 4-potential $\{\Delta v({\bf r}),\Delta{\bf B }(\bf r) \}$.
This happens if and only if $\Psi_0$ is an eigenstate of the operator \cite{Capelle2001}
\begin{equation} \label{NUcondition}
\Delta H = \sum_{j=1}^N \: [ \Delta v({\bf r}_j) - \Delta{\bf B}({\bf r}_j) \cdot  \vec{\sigma}_j] \:,
\end{equation}
where $\vec{\sigma}_j$ is the vector of Pauli matrices acting on the spin of the $j$th electron,
and we use units where the Bohr magneton $\mu_{\rm B}=1$. The entire spectrum $\{\Psi\}$ is invariant 
if $\Delta H$ is related to a constant of motion ({\em systematic} NU); all other cases
are called {\em accidental} NU \cite{Capelle2001}. 
Similar conditions for NU can be given for other multi-component generalizations of DFT, 
e.g. for current-carrying or superconducting systems \cite{Capelle2002}. Thus, NU appears to be 
a characteristic feature of generalized DFTs, and its fundamental and practical consequences need to be 
explored in detail. In the first place, it is important to know what types of NU can occur in 
practice. The following classes of examples have been identified in SDFT:

(a) $\{\Delta v =C,\Delta {\bf B}= B \hat{e}_z\}$,  with constant $B$, for systems with an energy gap 
and a collinear spin arrangement along the $z$-axis, where the $\{\Psi\}$ are eigenstates of $\hat{S}_z$ (systematic NU).
$B$ should be sufficiently small to avoid level crossings in order for $\Psi_0$ to remain the ground state.
As shown by Gidopoulos \cite{gidopoulos}, the mapping in the collinear case is unique
in the broader sense that spin-potentials $\{v_{\uparrow}({\bf r}),v_{\downarrow}({\bf r})\}$
which differ by more than a spin-dependent constant always have different ground states.

(b) In a fully spin polarized Kohn-Sham system  with $n_\uparrow({\bf r})=n({\bf r})$ and $n_\downarrow=0$ and an energy gap, 
there is an infinite number of spin-down Kohn-Sham potentials that produce the same ground state
(accidental NU).

(c) $\{\Delta v = \lambda u({\bf r}),\Delta {\bf B} = \lambda u({\bf r}) {\bf m}({\bf r})/m({\bf r}) \}$ 
for one-electron systems only, where $u({\bf r})$ is an arbitrary function and $\lambda$ is sufficiently small
(accidental NU) \cite{Barth1972}.

In Ref. \cite{Eschrig2001}, a general condition for NU in $N$-electron systems is given: If
$\Psi_0$ is invariant under the addition of a 4-potential, then this
4-potential must have the form
$\{\Delta v({\bf r}) = C, \Delta {\bf B}({\bf r}) = B \hat{e}({\bf r})\}$, i.e.,
$\Delta {\bf B}$ may possibly be noncollinear but must be constant in magnitude.
However, Argaman and Makov \cite{argaman2002} have raised doubts whether such noncollinear 4-potentials 
can really be found.

It thus appears that NU for $N>1$ electrons, while interesting and of potential practical relevance, is
limited to rather simple situations of collinear spin arrangement or full, ferromagnetic
spin polarization. The purpose of this paper is to examine NU in SDFT for arbitrary 
lattice systems. This has technical advantages over dealing with continuum systems since one can use linear algebra
methods in finite vector spaces. We discover two new, non-trivial classes 
of NU for $N$-electron systems with noncollinear spins, both of the ``accidental'' type.
These examples require the ground state 4-density to satisfy certain constraints on the lattice.
In the continuum limit, we show that only some well-known special cases survive, and we discuss 
consequences for practical applications.

We consider a noninteracting $N$-electron system on a finite-size lattice with $P$ lattice points 
whose specific geometry is not important for the following. We assume that the kinetic-energy operator $\hat{T}$
has been suitably discretized on this lattice, for example using a finite-difference approach.
The single-particle wave functions $\psi_j$ obey the following Schr\"odinger equation:
\begin{equation}\label{H0_eq}
[\hat{T} + \hat{V} - \hat{\bf B}\cdot \vec{\sigma}] \psi_j = E_j \psi_j \:,\qquad j=1,\ldots,2P\:.
\end{equation}
For the spatial part of the wave functions we use a localized basis, $\varphi_{ik}=\delta_{ik}$, $i,k=1,\ldots,P$.
The $j$th eigenstate on lattice site $k$ can then be written as
\begin{equation}
\psi_{jk} = \sum_{i=1}^P \varphi_{ik}( c_{ji}\alpha + d_{ji}\beta) =   c_{jk}\alpha + d_{jk}\beta\:,
\end{equation}
where $\alpha,\beta$ are the usual two-component spinors. The coefficients $c_{jk},d_{jk}$ follow
from diagonalizing the Hamiltonian matrix associated with the lattice-specific $\hat{T}$ and the given 
4-potential $\{v,{\bf B}\}$. The resulting 4-density on lattice site $k$ is
\begin{equation}\label{4-density}
\left(\begin{array}{c}n_k \\ m^x_k \\ m^y_k \\ m^z_k \end{array}\right)=
\sum_{j=1}^N \left( \begin{array}{c} |c_{jk}|^2 + |d_{jk}|^2\\
c_{jk}d_{jk}^* + c_{jk}^* d_{jk} \\ ic_{jk}d_{jk}^* - ic_{jk}^* d_{jk}\\
|c_{jk}|^2 - |d_{jk}|^2 \end{array}\right).
\end{equation}
A trivial case of NU arises when the lattice holds the maximum number of electrons allowed by the Pauli principle,
$N=2P$, for then $n=2$ and ${\bf m}=0$ for {\em all} external 4-potentials.
We now formulate the first new  nontrivial example for NU on lattice systems, which we call {\em local saturation} NU.

{\em {\bf Theorem I:} A noninteracting $N$-electron ground state on a $P$-point lattice is invariant under 
a perturbation with 4-potential $\{v_k',{\bf B}'_k = v_k' \: {\bf m}_k/m_k\}$, with arbitrary $v_k'$, 
that acts locally only on those lattice sites where $n_k = m_k$.}

Notice that for $N=1$ this reduces to the example (c) above, since 
in that case $n=m$ on all lattice sites. For $N>1$, it is not hard to show that the condition $n_k = m_k$ 
on a specific site $k$ is satisfied if and only if
\begin{equation}
\psi_{jk} =  f_j ( C_k \alpha + D_k \beta) \:,\qquad j=1,\ldots,N\:,
\end{equation} 
i.e., the lowest $N$ $\psi_j$ must have the same spin part, and therefore parallel magnetizations, on point $k$.
Let us now act on these states with a 4-potential that is nonzero on site $k$ only and vanishes on all other sites
$i\ne k$:
\begin{eqnarray}
\lefteqn{(v_k' - {\bf B}_k' \cdot \vec{\sigma})f_j ( C_k \alpha + D_k \beta)} \nonumber\\
&=& f_j [C_k v_k'  - D_k B_{x,k}' +iD_k B_{y,k}'  - C_k B_{z,k}'] \alpha \nonumber\\
&+& f_j[D_k v_k' - C_k B_{x,k}' -iC_k B_{y,k}'  + D_k B_{z,k}'] \beta  \:.
\end{eqnarray}
From this, it is straightforward to show that
\begin{equation}
v_k'[1 - ({\bf m}_k/n_k) \cdot \vec{\sigma}]f_j ( C_k \alpha + D_k \beta) = 0 
\end{equation}
for each $j$, which proves Theorem I. Local saturation NU can occur on one or more isolated lattice sites,
but also on groups of sites, which includes examples associated with the formation of ferromagnetic domains on the lattice.

The second new class of examples belongs to the Eschrig-Pickett type \cite{Eschrig2001} and will be referred to in
the following as {\em global noncollinear} NU.

{\em {\bf Theorem II:} A noninteracting $2$-electron ground state on a $P$-point lattice is invariant under a perturbation
with 4-potential $\{v'=0,{\bf B}' =  \lambda {\bf m}/m\}$,  $|\lambda|=const.$,  
if the ground state satisfies $\mbox{sign}(\lambda)(\bar{n}_{1} - \bar{n}_{2})/m = \mbox{const.}$}

Here, $\bar{n}_1$ and $\bar{n}_2$ denote the two occupied orbital densities, with $\bar{n}_{jk} = |c_{jk}|^2 + |d_{jk}|^2$,
$j=1,2$.
To prove Theorem II, one needs to show that the magnetic field ${\bf B}' =  \lambda {\bf m}/m$
causes at most an orthogonal rotation within the space spanned by the two lowest single-particle 
eigenstates, $\psi_1$ and $\psi_2$, which leaves the associated 2-particle Slater determinant invariant.
Thus,
\begin{equation} \label{GN1}
\left(\hat{H}_0 - \lambda\frac{{\bf m}\cdot \vec{\sigma}}{m} \right) (\gamma_{1i} \psi_1 + \gamma_{2i} \psi_2) = \varepsilon_i
(\gamma_{1i} \psi_1 + \gamma_{2i} \psi_2) \:, 
\end{equation}
for $i=1,2$, where the $\gamma_{ji}$ form an orthogonal $2\times 2$ matrix, and $\hat{H}_0$ is the
unperturbed single-particle Hamiltonian whose first two eigenstates and energies are $\psi_{1,2}$
and $E_{1,2}$, see Eq. (\ref{H0_eq}). Now consider a lattice site $k$.
After some straightforward algebra, using $m^\pm = m^x \pm im^y$, one arrives at the following expression:
\begin{widetext}
\begin{equation}\label{GN2}
\left( \begin{array}{cc} \displaystyle
\Big[c_{1k}E_1-\frac{\lambda_k}{m_k}(c_{1k}m^z_k + d_{1k}m^-_k) \Big] & \displaystyle
\Big[c_{2k}E_2-\frac{\lambda_k}{m_k}(c_{2k}m^z_k + d_{2k}m^-_k)\Big]\\[3mm] \displaystyle
\Big[d_{1k}E_1-\frac{\lambda_k}{m_k}(c_{1k}m^+_k - d_{1k}m^z_k)\Big] &\displaystyle
\Big[d_{2k}E_2-\frac{\lambda_k}{m_k}(c_{2k}m^+_k - d_{2k}m^z_k)\Big] \end{array}\right) 
\left(\begin{array}{c} \gamma_{1i} \\[3mm] \gamma_{2i}\end{array}\right)
=\varepsilon_i
\left(\begin{array}{cc}c_{1k} & c_{2k}\\[3mm] d_{1k} & d_{2k}\end{array}\right)
\left(\begin{array}{c} \gamma_{1i} \\[3mm] \gamma_{2i}\end{array}\right),
\end{equation}
which has the form of a generalized $2\times2$ eigenvalue problem. Eq. (\ref{GN2}) can be easily transformed
into a standard eigenvalue problem by multiplying with the inverse of the right-hand coefficient matrix. 
Using relations (\ref{4-density}), one finds after some manipulation
\begin{equation}
\left( \begin{array}{cc} \label{GN3}\displaystyle
\Big[-\frac{\lambda_k}{m_k}(\bar{n}_{1k}-\bar{n}_{2k}) + E_1 \Big]& \displaystyle
\Big[-2\frac{\lambda_k}{m_k}(c_{1k}^*c_{2k} + d_{1k}^*d_{2k})\Big]\\[3mm] \displaystyle
\Big[-2\frac{\lambda_k}{m_k}(c_{1k}c_{2k}^* + d_{1k}d_{2k}^*)\Big] & \displaystyle 
\Big[\frac{\lambda_k}{m_k}(\bar{n}_{1k}-\bar{n}_{2k}) + E_2 \Big]\end{array}\right)
\left(\begin{array}{c} \gamma_{1i} \\[3mm] \gamma_{2i}\end{array}\right)
=\varepsilon_i \left(\begin{array}{c} \gamma_{1i} \\[3mm] \gamma_{2i}\end{array}\right),
\end{equation}
\end{widetext}
which leads to a characteristic second-degree polynomial with solution
\begin{equation}
\varepsilon_{1,2} = \frac{E_1 + E_2}{2} \pm \sqrt{\frac{(\Delta E)^2}{4} + \lambda_k^2 - \lambda_k \Delta E \:
\frac{\bar{n}_{1k}-\bar{n}_{2k}}{m_k}} \:,
\end{equation}
where $\Delta E = E_1 - E_2$. We see immediately that for $\lambda=0$ this reduces to $\varepsilon_{1,2} = E_{1,2}$.

So far, the derivation was for a specific lattice site $k$. To ensure that the solution $\varepsilon_{1,2}$ and
the associated orthogonal eigenvectors $\gamma_{ji}$ are the same for all $P$ lattice sites, we need to impose
the constraints
\begin{equation}\label{GN4}
\mbox{sign}(\lambda_1)\:\frac{\bar{n}_{11}-\bar{n}_{21}}{m_1} = \ldots =
\mbox{sign}(\lambda_P)\:\frac{\bar{n}_{1P}-\bar{n}_{2P}}{m_P} \:,
\end{equation}
which completes the proof of Theorem II, and determines $\mbox{sign}(\lambda_k)$. 
Global noncollinear NU thus requires two-electron ground states 
whose orbital densities and total magnetization are related according to Eq. (\ref{GN4}). 
An explicit example for this will be given below.

Furthermore, from the normalization of the orbital densities, $\sum_{i=1}^P (\bar{n}_{1i}-\bar{n}_{2i})=0$, 
one finds that global noncollinear NU requires a total magnetization of the form
\begin{eqnarray}\label{MAG}
\mbox{sign}(\lambda_1)m_1 &=& -\mbox{sign}(\lambda_2)m_2 - \mbox{sign}(\lambda_3)m_3\nonumber\\
&& - \ldots - \mbox{sign}(\lambda_P)m_P \;.
\end{eqnarray}
Some additional remarks are in order:

{\em One-electron case.} For a single electron, (\ref{GN2}) reduces to 
\begin{equation}
\varepsilon = E_1 - \lambda_k n_k/m_k \:.
\end{equation}
Again, we require this to be the same on all lattice sites.
But, of course, $n=m$ everywhere for a single electron, so that we end up with
the condition $\lambda=const.$, i.e. $\varepsilon=E_1-\lambda$. This leads to the statement that 
any one-electron ground state is unchanged under the influence of 
a magnetic field ${\bf B}' =  \lambda{\bf m}/m$ (provided $\lambda$ is sufficiently small
such that the order of the lowest levels is not changed). This is
a special case already contained in the one-electron limit of Theorem I.

{\em $N$-electron case.}
Global noncollinear NU cannot occur for systems with more than two electrons, which
can be seen as follows. The 2-electron derivation is easily generalized up to the point where one arrives at 
a generalized eigenvalue problem similar to Eq. (\ref{GN2}), but of the type 
$\underline{R}\vec{\gamma} = \varepsilon \underline{S} \vec{\gamma}$ where $\underline{R},\underline{S}$ 
are $2\times N$ rectangular matrices, and $\vec{\gamma}$ is an $N$-component column vector.
Such underdetermined problems are singular, that is, one can
find at most two eigenvalues, all remaining $N-2$ eigenvalues are undefined \cite{SIAM}. 
Often one finds no eigenvalues at all.
This means that, except for trivial situations or by accident, there is no noncollinear field ${\bf B}'$ that results
only in a rotation within the single-particle ground-eigenspace. 
The $N$-particle ground-state Slater determinant is thus not invariant for $N>2$.

{\em 2-point lattices.} For the special case $P=2$ one can show that
{\em all} well-behaved 4-potentials produce two-electron ground states whose
magnetization has {\em same} magnitude on the two lattice sites, $m_1=m_2$.
This result is independent of electron interactions.
Global noncollinear NU is thus {\em always} present on 2-point lattices.

An interesting implication of this is that magnetizations with $m_1\ne m_2$ 
can arise on a 2-point lattice only as ensemble 4-densities of degenerate ground states.
Similar consequences of NU are expected for lattices with more than 2 points,
i.e., certain classes of 4-densities can only come from ensembles of degenerate ground states.
For the case of non-magnetic DFT, the topology of the $v$- and $n$-spaces on lattices
was recently clarified \cite{ullrich2002}, with the result that pure- and
ensemble-$v$-representable densities have the same mathematical measure. In SDFT,
this general statement no longer holds due to the much richer NU, as is evident from the 2-point
lattice example. 

\begin{figure}
\unitlength1cm
\begin{picture}(5.0,7.5)
\put(-6,-5.){\makebox(5.0,7.5){
\includegraphics{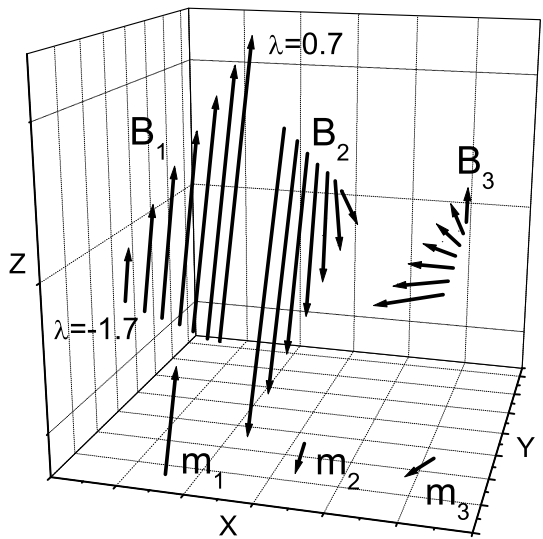}}}
\end{picture}
\caption{\label{figure1} Global noncollinear NU for a linear 3-point lattice
with $m_1 = m_2 + m_3$. All magnetic fields shown here, where ${\bf B}'_1 ={\bf B}_1+\lambda{\bf m}_1/m_1$ and
${\bf B}'_{2,3} ={\bf B}_{2,3}-\lambda{\bf m}_{2,3}/m_{2,3}$ (see Table \ref{table1}), produce the same 4-density
(keeping $v$ fixed).}
\end{figure}

Fig. \ref{figure1} illustrates an example for global noncollinear NU on a linear 3-point lattice 
with lattice constant $a$ and sites 1,2,3, using a finite-difference $\hat{T}$. To discover this and many other
examples, the 4-potential parameter space was numerically searched with a multidimensional simplex algorithm 
\cite{Nrecipes} until a two-electron ground state was found to satisfy Eq. (\ref{GN4}), with $m_1 = m_2 + m_3$, to within an
accuracy of $10^{-14}$ (similar numerical techniques yield examples for local saturation NU).
Measuring energies in units of $\hbar^2/2ma^2$, we give the 4-potential and resulting 4-density in Table \ref{table1}.
All magnetic fields ${\bf B}'= {\bf B} \pm \lambda {\bf m}/m$ ($+$ on site 1, $-$ on sites 2,3)
produce the {\em same} two-electron ground state 4-density (keeping $v$ fixed),
for $-1.7 < \lambda < 0.7$. Values of $\lambda$ outside that range
result in level crossings and thus different ground states.

\begin{table}
\caption{\label{table1} The 4-potential and 4-density used in Fig. \ref{figure1}, on lattice sites 1, 2, and 3.
Potentials and magnetic fields are measured in units of $\hbar^2/2ma^2$ (setting $\mu_{\rm B}=1$). The lattice 
4-density is dimensionless. Constraints (\ref{GN4}) and (\ref{MAG}) are satisfied, with $|\bar{n}_1-\bar{n}_2|/m=0.99756$
on each lattice site.}
\begin{ruledtabular}
\begin{tabular}{lccc}
 &  1  & 2 & 3  \\ \hline
$v$   &  -1.62192 & 1.55381 & 0.0 \\
$B^x$ &  0.87156 & -0.14000 & -0.50808 \\
$B^y$ & 0.15523 & 0.23990 & -0.69702 \\
$B^z$ & 1.76179 & -0.77994 & 0.11107 \\
$\bar{n}_1,\bar{n}_2$   & 0.96869,0.07157 & 0.02926,0.20498 & 0.00205,0.72345 \\
$m^x$ & 0.40883 & -0.10282 & -0.46681\\
$m^y$ & 0.08899 & -0.05986 & -0.54760 \\
$m^z$ & 0.79605 & -0.12989 & -0.07209\\
$m$   & 0.89931 & 0.17614 & 0.72317
\end{tabular}
\end{ruledtabular}
\end{table}

We now turn to the continuum limits of our lattice examples.

{\em 1. Local saturation NU.} It is possible that the local condition $n({\bf r}) = m({\bf r})$ is
satisfied  in lower-dimensional subspaces (e.g., points or lines) for a continuum system.
But 4-potentials that are confined to the same local subspaces
and vanish everywhere else are highly pathological (involving delta- or step
functions). Thus, only the well-known special cases of local saturation NU survive in the continuum limit, namely, the
1-electron and the completely polarized, ferromagnetic case.

{\em 2. Global noncollinear NU.} The number of constraints, see Eq. (\ref{GN4}), that need
to be imposed on ground states to exhibit this type of NU, grows with the number of
lattice sites. Thus, global noncollinear NU becomes increasingly rare for 
larger lattices, and is thus ruled out in the continuum limit, in agreement with Ref. \cite{argaman2002}. 
Again, only the 1-electron special case survives.

These findings are reassuring for the practical application of SDFT to electronic
structure calculations in atoms, molecules and solids. In the collinear case, all that
is required in a spin-dependent Kohn-Sham calculation is to fix two constants in the 
spin-up and spin-down channel, for example through the asymptotic behavior of the potentials. 
In the noncollinear case, a single constant appears to be sufficient.
In situations with full spin polarization, such as in half-metallic ferromagnets, 
the NU in SDFT is likely to result in the occurrence of discontinuities in the exchange-correlation potential,
which will require further study \cite{Capelle2001,Eschrig2001}.

On the other hand, spin systems on small lattices are of great interest in
the field of spintronics and quantum computation. From a basic point of view, for example, a quantum dot molecule
constitutes a two-point lattice. The results presented in this paper will be relevant for the manipulation 
of electronic charges and spins on such small 
lattice systems, for instance in performing qubit operations using external magnetic fields.

\acknowledgments
This work was supported in part by DOE Grant DE-FG02-04ER46151. The author
thanks Walter Kohn for suggesting this problem, and Giovanni Vignale
and Klaus Capelle for valuable discussions.


\end{document}